\documentclass[twocolumn,tighten]{aastex63}
\usepackage{verbatim}
\usepackage{color}
\usepackage{amsmath}
\usepackage{graphicx}
\usepackage{natbib}

\newcommand{\target}{GJ~1252}
\newcommand{\planet}{GJ~1252~b}

\newcommand{\rsun}{\ensuremath{R_\sun}}
\newcommand{\msun}{\ensuremath{M_\sun}}
\newcommand{\lsun}{\ensuremath{L_\sun}}
\newcommand{\rearth}{\ensuremath{R_\earth}}
\newcommand{\mearth}{\ensuremath{M_\earth}}

\newcommand{\rjup}{\ensuremath{R_{\rm Jup}}}
\newcommand{\mjup}{\ensuremath{M_{\rm Jup}}}
\newcommand{\teff}{\ensuremath{T_{\rm eff}}}

\newcommand{\feh}{[Fe/H]}

\newcommand{\kms}{\ensuremath{\rm km\,s^{-1}}}
\newcommand{\ms}{\ensuremath{\rm m\,s^{-1}}}
\newcommand{\rstar}{\ensuremath{R_s}}
\newcommand{\mstar}{\ensuremath{M_s}}
\newcommand{\lstar}{\ensuremath{L_s}}

\newcommand{\mic}{\ensuremath{\mu \rm m}}

\newcommand{\tess}{{\it TESS}}

\newcommand{\sig}[1]{\ensuremath{#1\sigma}}
\newcommand{\figr}[1]{Figure~\ref{fig:#1}}
\newcommand{\secr}[1]{Section~\ref{sec:#1}}

\newcommand{\tabr}[1]{\mbox{Table~\ref{tab:#1}}}

\newcommand{\msd}{\ensuremath{\rm m\,s^{-1}\,d^{-1}}}

\newcommand{\C}{\ensuremath{^{\circ}C\;}}

\shorttitle{GJ~1252~\lowercase{b}}
\shortauthors{Shporer et~al.}

\begin{document}

\title{\planet: A 1.2 \rearth\ planet transiting an M3-dwarf at 20.4 pc}


\author[0000-0002-1836-3120]{Avi Shporer}
\affil{Department of Physics and Kavli Institute for Astrophysics and Space Research, Massachusetts Institute of Technology, Cambridge, MA 02139, USA}

\author[0000-0001-6588-9574]{Karen A.~Collins}
\affil{Center for Astrophysics \textbar \ Harvard \& Smithsonian, 60 Garden Street, Cambridge, MA 02138, USA}

\author[0000-0002-8462-515X]{Nicola Astudillo-Defru}
\affil{Departamento de Matem\'atica y F\'isica Aplicadas, Universidad Cat\'olica de la Sant\'isima Concepci\'on, Alonso de Rivera 2850, Concepci\'on, Chile}

\author{Jonathan Irwin}
\affil{Center for Astrophysics \textbar \ Harvard \& Smithsonian, 60 Garden Street, Cambridge, MA 02138, USA}

\author{Xavier Bonfils}
\affil{Universit\'e Grenoble Alpes, CNRS, IPAG, F-38000 Grenoble, France}

\author[0000-0003-2781-3207]{Kevin I.~Collins}
\affil{George Mason University, 4400 University Drive, Fairfax, VA, 22030 USA}

\author{Elisabeth Matthews}
\affil{Department of Physics and Kavli Institute for Astrophysics and Space Research, Massachusetts Institute of Technology, Cambridge, MA 02139, USA}

\author[0000-0001-6031-9513]{Jennifer G.~Winters}
\affil{Center for Astrophysics \textbar \ Harvard \& Smithsonian, 60 Garden Street, Cambridge, MA 02138, USA}


\author{David R.~Anderson}
\affil{Astrophysics Group, Keele University, Staffordshire, ST5 5BG, UK}
\affil{Department of Physics, University of Warwick, Gibbet Hill Road, Coventry CV4 7AL, UK}
\affil{Centre for Exoplanets and Habitability, University of Warwick, Gibbet Hill Road, Coventry CV4 7AL, UK}

\author{James D.~Armstrong}
\affil{Institute for Astronomy, University of Hawaii, Maui, HI 96768, USA}

\author{David Charbonneau}
\affil{Center for Astrophysics \textbar \ Harvard \& Smithsonian, 60 Garden Street, Cambridge, MA 02138, USA}

\author[0000-0001-5383-9393]{Ryan Cloutier}
\affil{Center for Astrophysics \textbar \ Harvard \& Smithsonian, 60 Garden Street, Cambridge, MA 02138, USA}

\author[0000-0002-6939-9211]{Tansu Daylan}
\affil{Department of Physics and Kavli Institute for Astrophysics and Space Research, Massachusetts Institute of Technology, Cambridge, MA 02139, USA}
\affil{Kavli Fellow}

\author{Tianjun~Gan} 
\affil{Department of Astronomy and Tsinghua Centre for Astrophysics, Tsinghua University, Beijing 100084, China}

\author{Maximilian N.~G{\"u}nther}
\affil{Department of Physics and Kavli Institute for Astrophysics and Space Research, Massachusetts Institute of Technology, Cambridge, MA 02139, USA}
\affil{Juan Carlos Torres Fellow}

\author{Coel Hellier}
\affil{Astrophysics Group, Keele University, Staffordshire, ST5 5BG, UK}

\author{Keith Horne}
\affil{SUPA Physics and Astronomy, University of St. Andrews, Fife, KY16 9SS, Scotland, UK}

\author{Chelsea X.~Huang}
\affil{Department of Physics and Kavli Institute for Astrophysics and Space Research, Massachusetts Institute of Technology, Cambridge, MA 02139, USA}
\affil{Juan Carlos Torres Fellow}

\author[0000-0002-4625-7333]{Eric L.~N.~Jensen}
\affil{Department of Physics \& Astronomy, Swarthmore College, Swarthmore PA 19081, USA}

\author[0000-0003-0497-2651]{John Kielkopf}
\affil{Department of Physics and Astronomy, University of Louisville, Louisville, KY 40292, USA}

\author{Enric Palle}
\affil{Instituto de Astrofísica de Canarias, V\'{i}a L\'{a}ctea s/n, E-38205 La Laguna, Tenerife, Spain}
\affil{Departamento de Astrof\'{i}sica, Universidad de La Laguna, Spain}

\author{Ramotholo Sefako} 
\affil{South African Astronomical Observatory, P.O.~Box 9, Observatory, Cape Town 7935, South Africa}

\author[0000-0002-3481-9052]{Keivan G.~Stassun}
\affil{Vanderbilt University, Department of Physics \& Astronomy, 6301 Stevenson Center Lane, Nashville, TN 37235, USA}
\affil{Fisk University, Department of Physics, 1000 17th Avenue N., Nashville, TN 37208, USA}

\author[0000-0001-5603-6895]{Thiam-Guan Tan}
\affil{Perth Exoplanet Survey Telescope}
 
\author[0000-0001-7246-5438]{Andrew~Vanderburg}
\affil{Department of Astronomy, The University of Texas at Austin, Austin, TX 78712, USA}
\affiliation{NASA Sagan Fellow}
 


\author{George R.~Ricker}
\affil{Department of Physics and Kavli Institute for Astrophysics and Space Research, Massachusetts Institute of Technology, Cambridge, MA 02139, USA}

\author{David W.~Latham}
\affil{Center for Astrophysics \textbar \ Harvard \& Smithsonian, 60 Garden Street, Cambridge, MA 02138, USA}

\author{Roland Vanderspek}
\affil{Department of Physics and Kavli Institute for Astrophysics and Space Research, Massachusetts Institute of Technology, Cambridge, MA 02139, USA}

\author[0000-0002-6892-6948]{Sara Seager}
\affil{Department of Physics and Kavli Institute for Astrophysics and Space Research, Massachusetts Institute of Technology, Cambridge, MA 02139, USA}
\affil{Department of Earth, Atmospheric, and Planetary Sciences, Massachusetts Institute of Technology, Cambridge, MA 02139, USA}
\affil{Department of Aeronautics and Astronautics, MIT, 77 Massachusetts Avenue, Cambridge, MA 02139, USA}

\author{Joshua N.~Winn}
\affil{Department of Astrophysical Sciences, Princeton University, Princeton, NJ 08544, USA}

\author{Jon M.~Jenkins}
\affil{NASA Ames Research Center, Moffett Field, CA 94035, USA},




\author[0000-0001-8020-7121]{Knicole Colon}
\affil{NASA Goddard Space Flight Center, Exoplanets and Stellar Astrophysics Laboratory (Code 667), Greenbelt, MD 20771, USA}	

\author{Courtney D.~Dressing}
\affil{Department of Astronomy, University of California Berkeley, Berkeley, CA 94720-3411, USA}

\author{S\'ebastien L\'epine}
\affil{Department of Physics and Astronomy, Georgia State University, Atlanta, GA 30302-4106, USA}


\author[0000-0002-0638-8822]{Philip S.~Muirhead}
\affil{Department of Astronomy, Institute for Astrophysical Research, Boston University, 725 Commonwealth Avenue, Boston, MA 02215, USA}
 
\author[0000-0003-4724-745X]{Mark E.~Rose}
\affil{NASA Ames Research Center, Moffett Field, CA 94035, USA},
 
\author{Joseph D.~Twicken}
\affil{NASA Ames Research Center, Moffett Field, CA 94035, USA},
\affil{SETI Institute, Moffett Field, CA 94035, USA}
 
\author{Jesus Noel Villasenor}
\affil{Department of Physics and Kavli Institute for Astrophysics and Space Research, Massachusetts Institute of Technology, Cambridge, MA 02139, USA}


\begin{abstract}

We report the discovery of \planet, a planet with a radius of 1.193 $\pm$ 0.074~\rearth\ and an orbital period of 0.52~days around an M3-type star (0.381 $\pm$ 0.019~\msun, 0.391 $\pm$ 0.020~\rsun) located 20.385 $\pm$ 0.019~pc away. We use \tess\ data, ground-based photometry and spectroscopy, Gaia astrometry, and high angular resolution imaging to show that the transit signal seen in the \tess\ data must originate from a transiting planet. We do so by ruling out all false positive scenarios that attempt to explain the transit signal as originating from an eclipsing stellar binary. 
Precise Doppler monitoring also leads to a tentative mass measurement of 2.09 $\pm$ 0.56~\mearth.
The host star proximity, brightness ($V$ = 12.19 mag, $K$ = 7.92 mag), low stellar activity, and the system's short orbital period make this planet an attractive target for detailed characterization, including precise mass measurement, looking for other objects in the system, and planet atmosphere characterization. 

\end{abstract}

\section{Introduction}
\label{sec:intro}

The field of exoplanets has come a long way since the first discoveries at the end of the 20th century \citep{latham89, wolszczan92, mayor95}. One of the current frontiers in the study of exoplanets is that of small planets, smaller than Neptune and Uranus. The \textit{Kepler} mission led to the discovery of thousands of small planets \citep[e.g.,][]{borucki11, thompson18}. This in turn led to the measurement of the planet radius distribution, showing that within 1 au of Sun-like stars, small planets are more frequent than large (gas giant) planets \citep[e.g.,][]{borucki11, thompson18} and that there is a deficit (or local minimum) of planets with 1.5--2.0 \rearth\ \citep{fulton17}. However, the number of small planets with a well measured mass is still small, especially for planets with radius smaller than 2 \rearth. In addition, only for a few small planets has it been possible to characterize the atmosphere \citep[e.g.,][]{benneke19} or measure the stellar obliquity \citep[e.g.,][]{sanchis12, hirano12, albrecht13, sanchis15}.

The study of small planets is hampered by the lack of small planets 
orbiting stars that are bright enough for detailed follow-up investigations. The \tess\ mission \citep{ricker14,ricker15} is designed to overcome this problem by detecting transiting planet candidates orbiting bright stars positioned across almost the entire sky. Among those, planet candidates orbiting nearby M dwarf stars, at a distance of a few tens of parsecs, present a special opportunity, as their typical high proper motion and small size make it easier to rule out false positive scenarios \citep[e.g.,][]{crossfield19, vanderspek19}. This quickly clears the way for follow-up studies including mass measurement and atmospheric characterization. In addition, for nearby M dwarfs astrometric data from the Gaia mission will be sensitive to stellar and sub-stellar companions within 1 au \citep{perryman14, sozzetti14}.

Here we present the discovery of \planet, a small planet orbiting an M3-type star. The planet was initially discovered as a transiting planet candidate using \tess\ data. Based on the \tess\ data and additional follow-up data we are able to reject all false positive scenarios, showing it is a real planet. In addition, we were able to obtain a marginal mass measurement. Observations and data analysis of \tess\ data and ground-based photometry, spectroscopy, and high angular resolution imaging are described in \secr{obs}. Host star characterization is described in \secr{star}, and in \secr{fps} we go through all false positive scenarios showing that they are all rejected. In \secr{orbit} we investigate the radial velocity (RV) time series and search for an orbital RV signal. We discuss the newly discovered star-planet system in \secr{dis} and conclude with a brief summary in \secr{sum}.

\section{Observations and data analysis}
\label{sec:obs}

\subsection{\tess\ data}
\label{sec:tess}

\target\ was observed by Camera 2 of the \tess\ spacecraft during Sector 13 campaign, from 2019 June 19 to 2019 July 17. Listed in the \tess\ input catalog (TIC; \citealt{stassun18a}) as TIC 370133522 it was observed with a 2-minute cadence using an 11$\times$11 pixel subarray centered on the target\footnote{This target is part of Guest Investigator program 11180, PI: Courtney Dressing.}. The photometric data were processed through the Science Processing Operations Center (SPOC) pipeline \citep{jenkins16}, largely based on the predecessor \textit{Kepler} mission pipeline \citep{jenkins17}. 

The SPOC analysis had identified a transit-like signal in the target's light curve with a brief decrease in brightness of about 850 ppm (parts per million) every 0.518 days. Upon further inspection at the \tess\ Science Office it was added to the list of \tess\ objects of interest (TOIs) as TOI 1078.01.  We list astrometric and photometric information about the target in \tabr{info}.

We downloaded the \tess\ light curve from MAST\footnote{\url{https://mast.stsci.edu/}} and removed all flux measurements where the quality flag was set. This step removed 897 measurements out of the total of 20,479, or 4.4 \% of the data. We then proceeded with fitting the Presearch Data Conditioning (PDC) light curve \citep{smith12, stumpe14}. PDC light curves are corrected for contamination from nearby stars and instrumental systematics originating from, for example, pointing drifts, focus changes, and thermal transients. The instrumental trends are identified in a carefully selected sample of quiet stars located on the same CCD as the target and showing high correlation between each other \citep{smith12}. We note that for \target\ the uncertainty of individual 2-minute flux measurements is 16.4\% larger in the PDC light curves than in the raw photometry. This increased uncertainty accounts for potentially injected noise during the PDC process. 

We used \textit{Allesfitter}\footnote{\url{https://allesfitter.readthedocs.io}} \citep{gunther19a} for fitting a transit model. The \textit{Allesfitter} code uses \textit{ellc} \citep{maxted16} for the transit light curve model, \textit{emcee} \citep{foreman13} for sampling the multi dimensional parameter space and producing a posteriori distributions of fitted parameters, and a Gaussian Process \citep[GP,][]{foreman17} for modeling the correlated noise.

For the GP kernel we adopted the Mat\'ern-3/2 covariance\footnote{The covariance between two measurements with a time difference $t$ is $\sigma^2(1+\sqrt{3}t/\rho)\exp(-\sqrt{3}t/\rho)$} following \cite{rasmussen06} and \cite{foreman17}. This covariance function has two parameters: A characteristic amplitude $\sigma$ and time scale $\rho$, which we fitted while using a log-uniform prior. Before fitting the in-transit data we used the out-of-transit flux measurements to fit only the two noise model parameters. We then used the fitted values for those parameters as Gaussian priors when fitting a transit model to the in-transit data. This approach is designed to control the GP and prevent it from fitting also the astrophysical signal in addition to the noise. We confirmed that the fitted kernel time scale parameter ($\rho$) corresponds to a time scale that is much longer than the 2-minute cadence of the data, by about two orders of magnitude. The latter check is done to make sure the GP kernel does not over fit the data by attempting to fit the white noise. 

\begin{figure}
\includegraphics[width=3.6in]{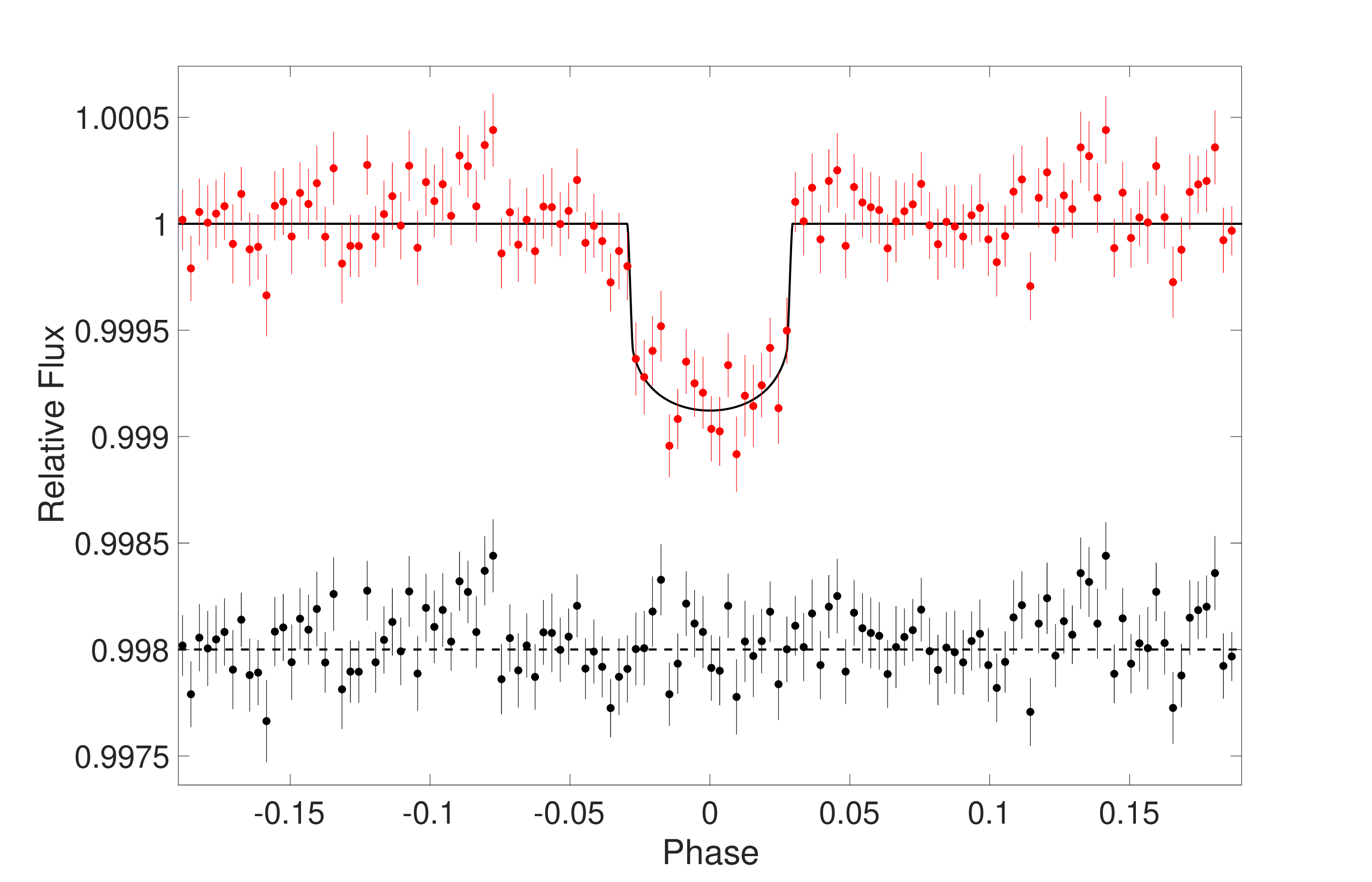}
\caption{Phase folded and binned \tess\ light curve (red). The fitted transit model is overplotted by a solid black line. The residual (data minus model) are shown at the bottom of the panel (in black), where a horizontal dashed black line is plotted for reference.
}
\label{fig:trlc}
\end{figure}

We fitted a transit light curve model while assuming a circular orbit and fitting seven free parameters with uniform priors: Orbital period $P$, specific mid transit time $T_0$ (which was chosen to be in the middle of the \tess\ time coverage, to minimize the covariance with $P$), planet to star radii ratio $R_p/R_s$, sum of star and planet radii divided by the orbital semi-major axis $(R_s + R_p)/a$, cosine of the orbital inclination $\cos i$, and two transformed quadratic limb-darkening law coefficients $q_1$ and $q_2$ (following \citealt{kipping13}). We used a Markov Chain Monte Carlo fitting procedure with 100 walkers with 10,000 steps each, and disregard the first 2,000 (20\%) of the steps. The number of steps was 30--50 times longer than the auto-correlation length of each of the fitted parameter chains. The fitted parameters are listed in \tabr{params} along with several parameters derived from the fitted parameters, and the transit light curve with the fitted model is plotted in \figr{trlc}. The derived parameters listed in \tabr{params} are based on the fitted parameters posterior distribution and the stellar parameters value and uncertainty (derived in \secr{star}).

In our analysis we assumed the contribution from other stars to the total flux in the photometric aperture was 2.2\% of the flux of \target, as reported in the TIC. While we did not assume an uncertainty on that parameter, it is three times smaller than the uncertainty on the measured transit depth. 

We tested our results for the fitted parameters by applying several variants of our model fitting, including:
\begin{itemize}
\item Fixing the limb darkening parameters to theoretical values of $q_1 = 0.36$ and $q_2 = 0.15$, based on \cite{claret17}.
\item Using Gaussian priors on $q_1$ and $q_2$ centered on the values above and with 0.1 standard deviation. This approach is used as an intermediate approach between fixing the limb-darkening coefficients to theoretical values and allowing them to vary freely with a uniform prior. The value of 0.1 for the standard deviation was chosen somewhat arbitrarily.
\item Fixing $q_1$ to 0.36 and allowing $q_2$ to vary freely with a uniform prior.
\item Repeating the analysis without GP, including the variants above.
\end{itemize}
All variants of the original analysis resulted in fully consistent results for the fitted parameters at the \sig{0.2} level. 

In a subsequent analysis we refined the ephemeris by simultaneously fitting \tess\ and ground-based light curves (described below in \secr{phot}). This was done while fitting only for $P$ and $T_0$ and using Gaussian priors on the rest of the parameters following their fitted values as derived when fitting only the \tess\ data. Those refined $P$ and $T_0$ values are listed in \tabr{params}.


\begin{deluxetable}{lcc}
\tablewidth{0pc}
\tabletypesize{\small}
\tablecaption{
    Target Information
    \label{tab:info}
}
\tablehead{
    \multicolumn{1}{c}{Parameter} &
    \multicolumn{1}{c}{Value}    &
    \multicolumn{1}{c}{Source}    \\
}
\startdata
TIC & 370133522 & TIC V8$^a$\\
R.A. & $\ \ \,$20:27:42.081 & Gaia DR2$^b$ \\
Dec. & -56:27:25.16 & Gaia DR2$^b$ \\
$\mu_{ra}$ (mas yr$^{-1}$)  &  $\ \ \,$  424.414 $\pm$ 0.074 & Gaia DR2$^b$ \\
$\mu_{dec}$ (mas yr$^{-1}$) & -1,230.623 $\pm$ 0.073 & Gaia DR2$^b$ \\
Parallax (mas) & 49.056 $\pm$ 0.046 & Gaia DR2$^b$ \\
Distance (pc)  & 20.385 $\pm$ 0.019 & Gaia DR2$^b$ \\
Epoch & 2015.5 & Gaia DR2$^b$\\
$B$ (mag)    & 13.655 $\pm$ 0.029 & AAVSO DR9$^c$ \\
$V$ (mag)    & 12.193 $\pm$ 0.056 & AAVSO DR9$^c$ \\
$Gaia$ (mag) & 11.2364 $\pm$ 0.0008 & Gaia DR2$^b$ \\
\tess\ (mag) & 10.1165 $\pm$ 0.0073 & TIC V8$^a$\\
$J$ (mag)    & 8.697 $\pm$ 0.019 & 2MASS$^d$ \\
$H$ (mag)    & 8.161 $\pm$ 0.034 & 2MASS$^d$ \\
$K$ (mag)    & 7.915 $\pm$ 0.023 & 2MASS$^d$ \\
\enddata
\tablenotetext{a}{\cite{stassun18a}.}
\tablenotetext{b}{\cite{gaia18}.}
\tablenotetext{c}{\cite{henden16}.}
\tablenotetext{d}{\cite{cutri03}.}
\end{deluxetable}

\subsection{Ground-based photometry}
\label{sec:phot}

\subsubsection{Las Cumbres Observatory (LCO)}
\label{sec:lco}

Five full transits of \target\ were observed using 1~m telescopes of the Las Cumbres Observatory (LCO\footnote{\url{http://lco.global}}; formerly named LCOGT) network \citep{brown13}. 

Two transits were observed with the Pan-STARSS $z$ filter (which is 1,040 \AA\ wide centered at 8,700 \AA), on the nights of UT 2019 August 24 and UT 2019 August 25 from the South African Astronomical Observatory (SAAO) and Siding Spring Observatory (SSO), respectively. Both observations used 35\,s exposure times and a defocus of 1.0\,mm, resulting in images with typical full width at half max (FWHM) of $\approx 3\arcsec$. 
Three transits were observed with the SDSS-$g$ filter, on the nights of UT 2019 September 19, UT 2019 September 20, and UT 2019 October 9. The two observations in September were done from SAAO, and the third observation, in October, was done from the Cerro Tololo Inter-American Observatory (CTIO). These three observations used 70\,s exposure times and a defocus of 0.3\,mm, resulting in images with typical FWHMs of $\approx 2\farcs5$. 
We used the {\tt TESS Transit Finder}, which is a customised version of the {\tt Tapir} software package \citep{jensen13}, to schedule our transit observations. The telescopes are equipped with $4096\times4096$ LCO Sinistro cameras having a pixel scale of 0$\farcs$389 pixel$^{-1}$ resulting in a $26\farcm5\times26\farcm5$ field of view.

The images were calibrated by the standard LCO BANZAI pipeline and the photometric data were extracted using the {\tt AstroImageJ} ({\tt AIJ}) software package \citep{collins17}. Circular apertures with a radius of 16 pixels ($6\farcs2$) were used to extract differential photometry from the Pan-STARSS $z$ band images. Circular apertures with a radius of 9 pixels ($3\farcs5$) were used to extract differential photometry from the SDSS-$g$ images. The nearest star in the Gaia DR2 and TIC v8 catalogues is $16\arcsec$ to the South of \target\ at the epoch of the follow-up observations, so the photometric apertures are not contaminated with significant flux from known nearby stars. All five LCO light curves are plotted together in \figr{groundlc}.

In addition to extracting the target light curve we have also extracted and carefully examined the light curves of all nearby stars within $2\farcm5$ of the target. On none of the nearby stars did we detect variability that could explain the observed signal in the \tess\ data if some of the light from a nearby variable (eclipsing binary) star were entering the target's aperture.


\subsubsection{MEarth}
\label{sec:mearth}

\begin{figure}
\includegraphics[width=3.5in]{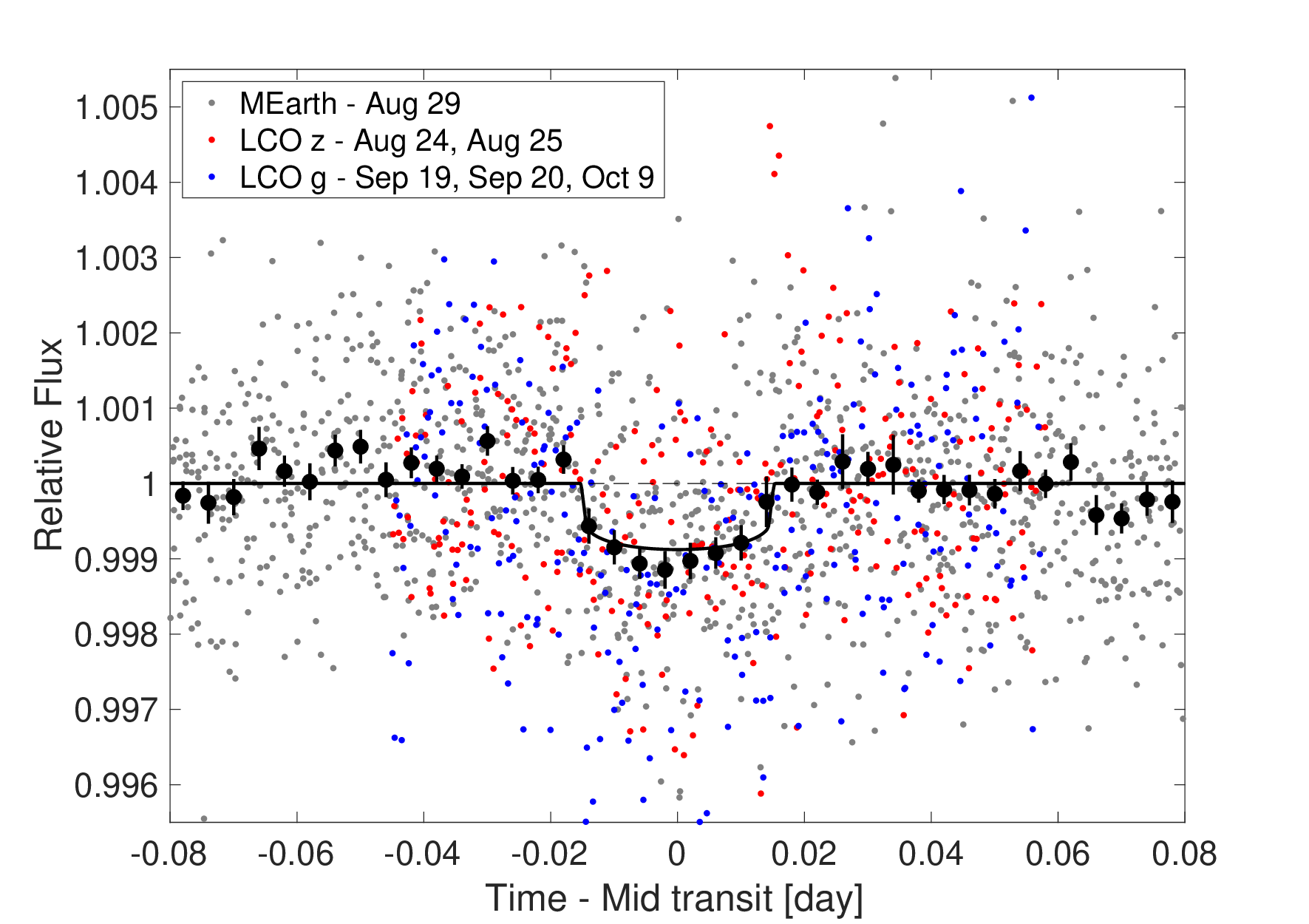}
\caption{Ground-based light curves of the target during predicted transit time. Two LCO 1m/Sinistro light curves obtained in the Pan-STARRS~z band are shown in red, three LCO 1m/Sinistro light curves obtained in the SDSS-g band are shown in blue, and a MEarth light curve is shown in gray. The legend lists the dates the light curves were obtained on. Binning all light curves results in the black points, and a horizontal dashed black line is marked at unit relative flux for reference. The binned light curve shows a transit light curve consistent with that seen in the \tess\ data. While the light curve shape is in principle sensitive to wavelength due to the stellar limb darkening wavelength dependency, the quality of the ground-based data is not sufficient for it to be sensitive to the small variations in shape between different bands.
}
\label{fig:groundlc}
\end{figure}

We used the MEarth array of seven 0.4~m telescopes \citep{nutzman08, irwin09}, located at CTIO, to observe \target\ on UT 2019 August 29 for 4.5 hours centred on the predicted transit time. MEarth uses the RG715 filter, with a wavelength range that is encompassed by the \tess\ band wavelength range. We used all seven telescopes to observe the target simultaneously with an exposure time of 60 seconds, while applying a defocus that brings the half flux diameter to 12 pixels. \figr{groundlc} shows the combined light curve from all seven MEarth telescopes. 

As with the LCO data, we examined the MEarth light curves of all nearby stars within 2.5 arcmin of the target. None of the nearby stars showed variability amplitude (or eclipse depth) that can explain the observed signal in the \tess\ data.\\


All ground-based light curves used here are available on the exoplanet follow-up observing program for \tess\ (ExoFOP-TESS\footnote{https://exofop.ipac.caltech.edu/tess/}).
The combined ground-based data, from both LCO and MEarth, plotted in \figr{groundlc}, show a shallow transit signal with a low signal-to-noise ratio. Although noisy, the observed signal is consistent with the transit seen in the \tess\ data. Therefore, as mentioned above, we used it to refine the transit ephemeris by simultaneously fitting \tess\ and ground-based light curves with $P$ and $T_0$ as the only free parameters. The fitted values for those parameters are listed in \tabr{params}. In this fitting we have adopted Gaussian priors on $R_p/R_s$, $(R_p + R_s)/a$, $\cos i$, and the two limb darkening parameters in the \tess\ band, following their fitted values as derived when fitting only the \tess\ data. We adopted Gaussian priors also for the limb darkening coefficients of the ground-based light curves, centered on theoretical values from \cite{claret12} and a \sig{1} value of 0.10. In a separate fit we allowed the limb darkening coefficients of the ground-based light curve to vary freely, resulting in identical values for $P$ and $T_0$.

\begin{deluxetable}{lll}
\tablewidth{0pc}
\tabletypesize{\small}
\tablecaption{
    Fitted and derived Parameters
    \label{tab:params}
}
\tablehead{
    \multicolumn{1}{c}{~~~~~~~~~~Parameter~~~~~~~~~~} &
    \multicolumn{1}{l}{Value}                     &
    \multicolumn{1}{l}{Uncertainty}    
}
\startdata
\multicolumn{3}{l}{\textit{Host star Parameters}} \\ 
\mstar\ (\msun) & 0.381 & 0.019 \\
\rstar\ (\rsun) & 0.391 & 0.020\\ 
\lstar\ (\lsun) &  0.0196 &  $_{-0.0023}^{+0.0026}$ \\ 
\teff\ (K)      & 3,458 & $_{-133}^{+140}$\\ 
\feh\           & 0.1 & 0.1 \\ \\
\multicolumn{3}{l}{\textit{Light curve fitted parameters\tablenotemark{a}}} \\ 
$R_p / R_s$         & 0.02802 & $_{-0.00097}^{+0.00090}$  \\ 
$(R_s + R_p) / a$   & 0.203   & $_{-0.011}^{+0.015}$      \\ 
$\cos{i}$           & 0.086   & $_{-0.036}^{+0.031}$  \\ 
$T_0$ (BJD - 2,458,000) & 668.09739   & $_{-0.00029}^{+0.00032}$ \\ 
$P$   $\mathrm{(d)}$    &    0.5182349 & $_{-0.0000050}^{+0.0000063}$   \\
$q_{1, \tess}$ & $0.23$ & $_{-0.16}^{+0.31}$ \\ 
$q_{2, \tess}$ & $0.36$ & $_{-0.25}^{+0.38}$  \\ 
$\ln \sigma$\tablenotemark{b} & -9.12 & 0.13 \\
$\ln \rho$\tablenotemark{b}   & -1.81 & $_{-1.03}^{+0.88}$ \\ \\
\multicolumn{3}{l}{\textit{RV orbit fitted parameters\tablenotemark{a}}} \\ 
$K$ (\ms)                       & 3.17   & 0.85 \\
$\gamma$ (\ms)                  & 7,483.56   & 0.70 \\
$\dot{\gamma}$ (\msd)           & -1.13  &  0.24\\
$\sigma_{RV}$ (\ms)             & 0.93   & $_{-0.58}^{+0.77}$\\ \\
\multicolumn{3}{l}{\textit{Derived Parameters}} \\ 
$i$ ($^{\circ}$)             & 85.0 & $_{-1.8}^{+2.1}$ \\
$a$ (AU)                     & 0.00916 & 0.00076 \\
$b$                          & 0.44  & $_{-0.17}^{+0.12}$ \\ 
$T_\mathrm{tot}$\tablenotemark{c} (h) & 0.734 & $_{-0.014}^{+0.018}$ \\ 
$T_\mathrm{full}$\tablenotemark{d}  (h) & 0.683 & $_{-0.016}^{+0.021}$ \\ 
$M_p$ (\mearth) & 2.09& 0.56\\
$R_p$ (\rearth) & 1.193 & 0.074\\
$T_\mathrm{eq}$\tablenotemark{e} (K) & 1089 & 69 
\enddata
\tablenotetext{a}{Assuming a circular orbit and adopting BJD of 2,458,751 as the reference epoch for the RV slope.}
\tablenotetext{b}{Noise model parameter of the Mat\'ern-3/2 GP kernel. Fitted using the out-of-transit data.}
\tablenotetext{c}{From 1st to last (4th) contacts.}
\tablenotetext{d}{From 2nd to 3rd contacts.}
\tablenotetext{e}{Assuming zero albedo and complete heat circulation between the day and night hemispheres.}
\end{deluxetable}

\subsection{High resolution spectroscopy}
\label{sec:spec}

To monitor the target's RV and measure the spectroscopic orbit we used the High Accuracy Radial velocity Planet Searcher (HARPS; \citealt{mayor03}). 
Our strategy was to acquire two consecutive RV points per night of 1,200 seconds each. We ended with a time series of 20 RV points spanning 11 days, between 2019 September 19 and September 30. To derive RVs we followed \citet{astudillo17}: RVs from HARPS Data Reduction Software were used to shift reduced spectra \citep{lovis07} to a common reference frame. A median spectrum was computed and shifted in several RV steps. From each step we derived the likelihood, whose maximum resulted in the RV used hereafter. The spectra signal-to-noise ratio (SNR) varies between 9--14 at 600 nm, equivalent to a RV precision of 5.0--1.8 \ms, with an average of 2.9 \ms. The RV dispersion is 4.7 \ms.

The HARPS RVs are listed in \tabr{rvs}. We also list in \tabr{rvs} five HARPS RVs derived from archival HARPS spectra that were available to us through the European Southern Observatory (ESO) online archive\footnote{\url{http://archive.eso.org/wdb/wdb/adp/phase3\_spectral/form}}. Those spectra were obtained during 2008 (two spectra) and 2011 (three spectra).



\begin{deluxetable*}{lcccccccc}
\tablewidth{0pc}
\tabletypesize{\small}
\tablecaption{
    Radial Velocities\tablenotemark{a}
    \label{tab:rvs}
}
\tablehead{
    \multicolumn{1}{c}{BJD} &
    \multicolumn{1}{c}{RV}    &
    \multicolumn{1}{c}{$\sigma_{RV}$}    &
    \multicolumn{1}{c}{$S$\tablenotemark{b}}    &
    \multicolumn{1}{c}{$\sigma_S$}    &
    \multicolumn{1}{c}{FWHM\tablenotemark{c}}    &
    \multicolumn{1}{c}{Contrast\tablenotemark{d}}    &
    \multicolumn{1}{c}{BIS\tablenotemark{e}}    &
    \multicolumn{1}{c}{SNR\tablenotemark{f}}    \\
    \multicolumn{1}{c}{} &
    \multicolumn{1}{c}{\ms}    &
    \multicolumn{1}{c}{\ms}    &
    \multicolumn{1}{c}{}    &
    \multicolumn{1}{c}{}    &
    \multicolumn{1}{c}{\kms}    &
    \multicolumn{1}{c}{}    &
    \multicolumn{1}{c}{\ms}    &
    \multicolumn{1}{c}{}   \\ 
}
\startdata
2454658.826649 & 7488.57 & 2.87 & 0.599 & 0.092 & 4.05 & 28.3 & -9.11 & 15.2 \\ 
2454660.877783 & 7489.37 & 2.19 & 1.116 & 0.190 & 4.08 & 28.5 & -31.26 &  9.3 \\ 
2455673.895142 & 7492.19 & 3.27 & 0.340 & 0.130 & 4.06 & 28.2 & -13.57 & 14.6 \\ 
2455801.570066 & 7485.14 & 2.11 & 0.239 & 0.218 & 4.07 & 28.4 & -6.96 & 10.3 \\ 
2455826.613565 & 7484.39 & 2.54 & 0.199 & 0.125 & 4.04 & 28.2 & -29.60 & 17.5 \\ 
2458745.658699 & 7488.57 & 2.87 & 0.556 & 0.175 & 4.10 & 27.9 &  8.23 & 19.1 \\ 
2458746.486546 & 7489.37 & 2.19 & 0.463 & 0.099 & 4.10 & 28.2 &  4.40 & 23.6 \\ 
2458746.712953 & 7492.19 & 3.27 & 0.551 & 0.205 & 4.08 & 27.9 &  6.17 & 17.0 \\ 
2458747.485652 & 7485.14 & 2.11 & 0.471 & 0.091 & 4.09 & 28.2 &  6.32 & 24.4 \\ 
2458747.500374 & 7484.39 & 2.54 & 0.371 & 0.123 & 4.11 & 28.2 & -0.07 & 20.9 \\ 
2458747.720416 & 7490.68 & 2.90 & 0.560 & 0.197 & 4.10 & 27.8 & -9.22 & 19.1 \\ 
2458747.735276 & 7490.77 & 2.80 & 0.564 & 0.191 & 4.10 & 27.7 &  13.03 & 19.7 \\ 
2458748.521492 & 7481.36 & 2.21 & 0.540 & 0.103 & 4.10 & 28.2 &  2.46 & 23.6 \\ 
2458748.715540 & 7488.54 & 2.62 & 0.451 & 0.183 & 4.09 & 27.7 &  5.22 & 21.1 \\ 
2458748.730424 & 7485.24 & 2.75 & 0.537 & 0.194 & 4.08 & 27.7 &  0.37 & 20.2 \\ 
2458750.599928 & 7481.88 & 1.81 & 0.454 & 0.087 & 4.09 & 28.0 &  8.77 & 28.2 \\ 
2458750.614788 & 7481.04 & 2.00 & 0.489 & 0.108 & 4.10 & 28.0 & -1.94 & 26.0 \\ 
2458752.704922 & 7476.67 & 2.40 & 0.522 & 0.178 & 4.09 & 27.8 & -3.64 & 22.8 \\ 
2458752.718624 & 7475.12 & 2.37 & 0.329 & 0.169 & 4.07 & 27.7 &  10.34 & 23.0 \\ 
2458754.574260 & 7483.96 & 3.25 & 0.692 & 0.176 & 4.12 & 28.1 &  3.93 & 17.1 \\ 
2458754.582512 & 7478.89 & 3.50 & 0.478 & 0.194 & 4.09 & 28.0 &  9.00 & 16.2 \\ 
2458755.570157 & 7482.90 & 2.91 & 0.595 & 0.191 & 4.09 & 27.9 &  6.85 & 19.0 \\ 
2458755.577934 & 7483.93 & 2.81 & 0.390 & 0.194 & 4.10 & 27.9 &  6.63 & 19.6 \\ 
2458756.592463 & 7481.27 & 4.84 & 0.797 & 0.356 & 4.09 & 27.7 &  8.39 & 12.5 \\ 
2458756.600575 & 7480.20 & 4.95 & 0.652 & 0.326 & 4.09 & 27.6 & -7.36 & 12.3 
\enddata
\tablenotetext{a}{The Gaia DR2 RV is 7.34 $\pm$ 0.33 \kms \citep{gaia18}.}
\tablenotetext{b}{Activity $S$-index, calculated following \citet{astudillo17}.}
\tablenotetext{c}{Spectral cross correlation fucntion (CCF) full width at half max.}
\tablenotetext{d}{CCF Contrast.}
\tablenotetext{e}{Bisector span.}
\tablenotetext{f}{Signal to noise ratio at HARPS order 60, at 612 nm.}
\end{deluxetable*}

\subsection{High angular resolution imaging}
\label{sec:imaging}

\begin{figure}
\includegraphics[width=\linewidth]{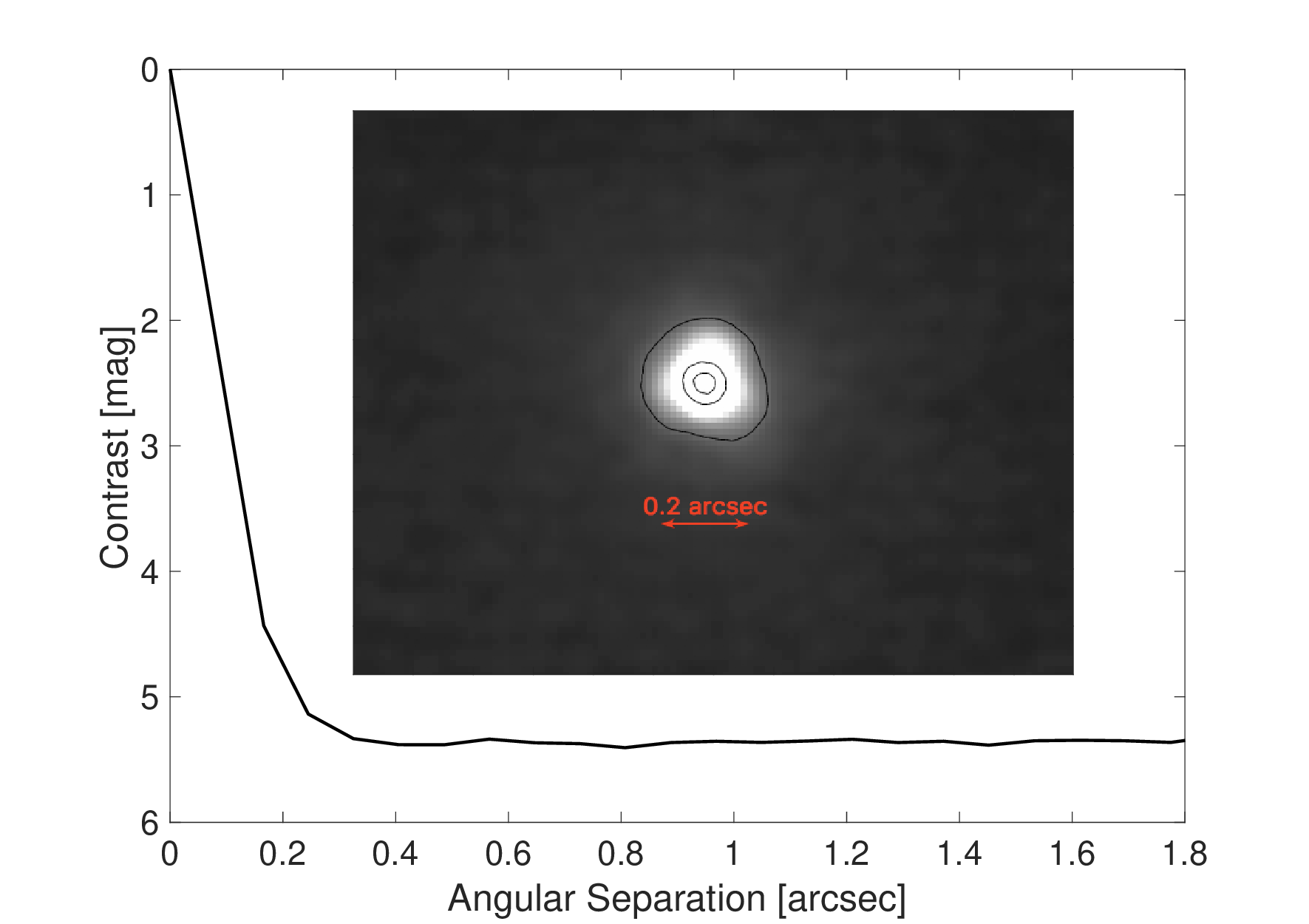}
\caption{VLT/NaCo adaptive optics image in Br$\gamma$ band (inset) and the derived contrast curve (solid black line). The inset also shows contours (in black lines) outlining the observed point spread function.
}
\label{fig:aoimage}
\end{figure}

The constant light from unresolved stars within the \tess\ photometric aperture can reduce the amplitude of the transit signal, thus reducing the inferred planet radius. They can even be the source of false positives if the companion itself is an eclipsing binary \citep{ciardi15}. We used adaptive optics imaging at VLT/NaCo to search for such visual companions. A total of 9 images were collected, each with 8s exposure, in the Br$\gamma$ band centered at 2.166 $\mu$m. The telescope was dithered between each exposure, to allow a sky background to be constructed from the science frames. We used a custom set of IDL codes to process the data following a standard process: bad pixels were removed, images were flat fielded and a sky background subtracted, the stellar position was aligned between frames, and the images were coadded. Our contrast sensitivity is calculated by inserting fake companions, and scaling their brightness until they are detected at 5$\sigma$. No companions are detected in the field of view. The image and the sensitivity curve are shown in Figure \ref{fig:aoimage}.

\section{Stellar Parameters}
\label{sec:star}

We estimated the stellar parameters using the empirical relations of \cite{mann15}. We first derived the absolute magnitude in $K$ band using the observed magnitude and Gaia parallax (while accounting for the systematic offset described by \citealt{stassun18b}), resulting in $M_K$ = 6.372 $\pm$ 0.023 mag.

Next we used the empirical relation between stellar mass and $M_K$ with the coefficients listed in  \citet[][see their Table 6 and Equation 2]{mann19}. Assuming a conservative uncertainty of 5\% this results in $M_s = 0.381 \pm 0.019\ \msun$. For comparison, we calculated the stellar mass with the empirical relation of \citet[see their Table 1 and Equation 10]{mann15}, and \citet[see their Table 13 and Equation 11]{benedict16}, resulting in $0.405 \pm 0.020\ \msun$ and $0.424 \pm 0.022\ \msun$, respectively. The latter two estimates are within 10\% or \sig{2} from the above estimates.

We estimated the stellar radius using the empirical relation between radius and $M_K$ derived by \citet[][see their Table 1]{mann15}, resulting in $R_s = 0.391 \pm 0.020\ \rsun$, assuming a conservative uncertainty of 5\%. For comparison, we estimated the stellar radius using the radius-mass empirical relation derived by \citet[][their Equation 10]{boyajian12}, resulting in $0.368 \pm 0.018\ \rsun$, which is 6\% or \sig{1.1} from the estimate above. 

To estimate the stellar effective temperature we first calculated the bolometric correction BC$_{K}$ to $M_K$. We do that using the empirical relations between BC$_{K}$ and the $V-J$ color given by \citet[][Table 3]{mann15}. We found BC$_{K}$ = 2.64 $\pm$ 0.13 mag and in turn a bolometric magnitude of $M_{\rm bol}$ = 9.01 $\pm$ 0.13 mag, which is equivalent to a bolometric luminosity of \lstar\ = $0.0196_{-0.0023}^{+0.0026}$ \lsun. Finally, using the Stefan-Boltzmann law we got \teff = $3458_{-133}^{+140}$ K. The stellar parameters derived here correspond to a spectral type of M2.5 \citep{pecaut13}\footnote{\url{http://www.pas.rochester.edu/$\sim$emamajek/\\EEM\_dwarf\_UBVIJHK\_colors\_Teff.txt}}.

We note that our derived stellar mass, radius, and temperature, are within \sig{1.5} of the values reported by \cite{muirhead18} and TIC V8 \citep{stassun18a}.

We estimated the stellar metallicity using the method described by \cite{dittmann16}. We identified stars in the \cite{dittmann16} sample with similar color and $M_K$ to \target, and calculated a weighted mean of the metallicity of those stars where the weights are the distance from the target position in the color-magnitude diagram. This resulted in a metallicity of \feh\ = 0.1~$\pm$~0.1.





Inspection of the HARPS spectra showed absorption in the $H_{\alpha}$ line. According to \cite{walkowicz09}, M3-type stars showing absorption in $H_{\alpha}$ are either moderately active or inactive, but not strongly active. This is consistent with the \cite{astudillo17} measurement of $\log R'_{HK} = -5.356 \pm 0.357$ based on the archival HARPS spectra, which according to their study of the correlation between $\log R'_{HK}$ and stellar rotation corresponds to a rotation period of 72 days.

\begin{figure}
\includegraphics[width=3.5in]{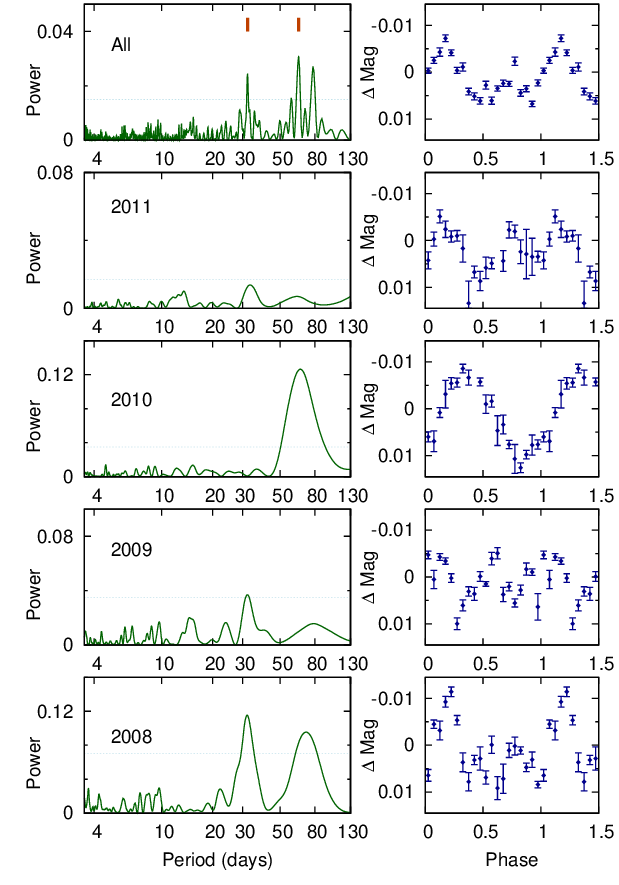}
\caption{Period analysis of WASP times-series photometry. The bottom four rows show the periodograms of four seasonal data sets (left panels) and the phase-folded and binned light curve using the strongest periodogram component (right panel). The top panels show the periodogram of all four seasons combined. The strongest peak, at 64.2 days, is marked in red, along with its first harmonic of that period at 32.1 days. The horizontal dashed blue line is the 1\% false alarm level estimated using the method of \citet{maxted11}.
}
\label{fig:wasp}
\end{figure}

We attempted to measure the stellar rotation period using multi-season time-series photometry from WASP-South (see \citealt{pollacco06}), taken between 2008 -- 2011. The period analysis of each observing season is shown in \figr{wasp}, along with the analysis of all seasons combined (top panel). The strongest period component in the combined analysis is at $64\pm4$ days, taking into account that the modulation may not be coherent over the full WASP dataset. This is close to the prediction above based on $\log R'_{HK}$.

\section{Rejecting false positive scenarios}
\label{sec:fps}

In the subsections below we consider the various false positive scenarios that might lead to a transit-like signal in the \tess\ data and show that they are rejected based on the data we have accumulated.

\subsection{The target is an EB}
\label{sec:eb}

The transit-like signal might be a grazing eclipse signal due to a stellar eclipsing companion. However, such massive companion would make the target show a large RV variation, which is not observed. We have a total of 25 HARPS RVs, five from archival spectra and 20 from spectra obtained as part of this work, listed in \tabr{rvs}. Those RVs were taken over 11 years (2008 -- 2019) and span less than 20 \ms\ with individual RV uncertainties of 2--5 \ms. We note that the replacement of the HARPS fiber in 2015 is expected to induce an RV zero point offset of only a few \ms\ for early-type M dwarfs \citep{locurto15}, so it should not affect the HARPS RV span significantly.
In addition, the Gaia DR2 RV is $7.34 \pm 0.33\ \kms$, with an RV uncertainty typical of single stars \citep{katz19}, and consistent with the HARPS RVs (see \tabr{rvs}).
This rules out a stellar companion since a 0.1 \msun\ binary companion at the transit period would induce an RV semi-amplitude of the target of 43 \kms. Moreover, a 1 \mjup\ companion would induce an RV semi-amplitude of 0.47 \kms\ which the observed RVs also reject.

Yet another way to rule out the transit companion being a massive object (massive planet or more massive object) is through orbital phase modulations in the \tess\ light curve. A massive orbiting companion is expected to induce detectable phase modulations along the orbit, as recently detected in the \tess\ light curve of a few star-planet systems (WASP-18, \citealt{shporer19}; KELT-9, \citealt{wong19}; WASP-121, \citealt{daylan19}; for a review see \citealt{shporer17}). A 10 \mjup\ companion orbiting \target\ at the transit period is expected to induce modulations with a semi-amplitude of $\approx$250 ppm. We have tested the detectability of such a signal through injection and recovery. We injected a phase curve modulation signal expected to be induced by a 10 \mjup\ companion into the light curve after removing the transit signal. A periodogram of that light curve showed a peak at the expected period that was ten times higher than any other peak which is not a harmonic of the orbital period. We concluded that massive planets, or more massive objects, at the transit period would have induced a clear orbital phase modulation, which is not observed in the \tess\ data.

\subsection{Nearby EB and background EB}
\label{sec:beb}

\begin{figure*}
\begin{center}
\includegraphics[width=7.0in]{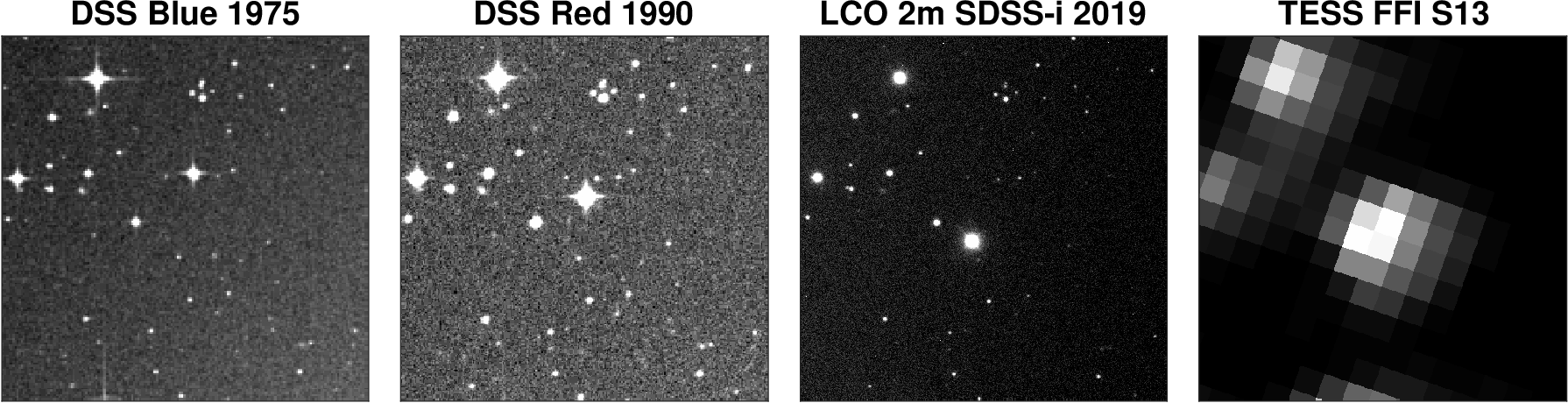}
\end{center}
\caption{The three left panels show the target's field of view in Digital Sky Survey images taken in 1975 (leftmost panel) and 1990 (second panel from the left), and in an LCO 2m Faulkes Telescope South image taken in 2019 (second panel from the right). The motion of the target on the sky between the images is clearly seen. The rightmost panel shows the target in one of the \tess\ FFIs during Sector 13 (also taken during 2019). In all four panels North is up and East is to the left, and they show a 5 $\times$ 5 arcmin field of view centered on the target's position at epoch 2015.5.
}
\label{fig:fcharts}
\end{figure*}

The transit signal in the \tess\ data can be originating from an eclipsing binary that is not associated with the target but whose light is blended with the target in the \tess\ PSF, which is roughly an arc-minute wide (see \figr{fcharts}).

To test that scenario we observed the target using ground-based seeing-limited facilities (see \secr{phot}). Those observations show that nearby stars that are resolved in our ground-based observations do not show variability, and the photometric precision for each of the stars is sufficient to rule out an eclipse deep enough to induce the observed transit on the target in the \tess\ data, given the brightness difference in \tess\ magnitude. Moreover, we have identified in \figr{groundlc} a transit-like feature in the target light curve in data obtained by two different ground-based facilities. 

It is also possible that there is an EB in the target's background that is blended with the target in the ground-based seeing-limited data. That scenario is ruled out with the help of the star's high proper motion. Archival images from 1975 and 1990 offer an unobstructed view toward the current position of \target. Those images (see \figr{fcharts}) rule out any background stars brighter than $R\approx19.1$ mag, which is good enough to exclude any eclipsing binaries capable of producing a transit-like signal in the \tess\ data with the observed amplitude.


\subsection{A gravitationally bound EB}
\label{sec:beb}

The transit-like feature might originate from another star that is gravitationally bound to the target, such that it has the same proper motion and is not resolved in ground-based observations, and if that other star is itself an eclipsing binary or a transiting star-planet system.

Visual examination of the transit light curve shape (see \figr{trlc}) shows that it has a short ingress and egress and most of the transit is spent in the so called ``flat bottom" part (2nd to 3rd contacts). This is expressed quantitatively by the measured flat bottom duration (see \tabr{params}) to be $93.4 \pm 3.6\ \%$ of the total transit duration. Therefore, the relative duration of the ingress and egress is $< 9$ \% (\sig{3}). An ingress and egress duration of 9 \% of the total transit duration is also rejected by the data, as the residuals of such a model show significantly increased scatter (by 5\%) and systematic features during ingress and egress. The reduced $\chi^2$ increased by 6\%, which is significant given the number of degrees of freedom. 

The upper limit on the ingress and egress duration is also an upper limit on the possible radii ratio that can produce the observed light curve, whether it is on the target star or originating from a fully blended gravitationally bound star which is itself an EB or star-planet system. 
Therefore the transiting or eclipsing object must be smaller than 0.34 \rjup\ which is close to the radius of Neptune. Hence the transiting object cannot be a star, a brown dwarf, or a gas giant planet. 

\subsection{A gravitationally bound transiting star-planet system}
\label{sec:btr}

We are left with the possibility that a gravitationally bound star has a transiting planet. From the above the planet-to-star radii ratio must be below 9\%, making the transit depth no more than 0.8\%. In order to produce the observed transit depth in the \tess\ data of 0.09\% the gravitationally bound star must contribute at least 9\% to the total light in the \tess\ band. Given the contrast curve measured by AO imaging (see \secr{imaging} and \figr{aoimage}) this means the bound star must be not fainter than 2.6 mag below the target and thus within 0.1 arcsec from the target, since the AO imaging was done in Br $\gamma$ band, where the brightness difference between the target and a smaller M dwarf companion will be smaller than in the optical (\tess\ band).

The stellar mean density based on the characteristics of the transit signal is 9.74 $\pm$ 0.65 g cm$^{-3}$ (assuming the transiting object is a planet, with a negligible mass compared to the host star). This is consistent with the mean density given the estimates of the stellar radius and mass, of 9.00 $\pm$ 1.45 g cm$^{-3}$. However, smaller stars have larger mean density. A 0.3 \msun\ main sequence star has a mean density of about 15 g cm$^{-3}$, which is inconsistent with the transit observation. While this rules out a transit on a bound star with mass below 0.3 \msun, a star more massive than 0.3 \msun\ and still of lower mass than the target (of 0.38 \msun) should have been bright enough to be detected in the HARPS spectra, which show only lines from the target (i.e.~it is single lined). Therefore we can rule out the transit signal in the \tess\ data as originating from a planet transiting a bound star.

Another argument against this scenario is the following. An angular distance of 0.1 arcsec corresponds to a sky-projected physical distance of 2.0 au. Since the 11 years coverage of the HARPS RVs (2008 -- 2019) show constant RV to well within 100 \ms\ we can rule out low-mass companions at that distance down to a planet mass. For example, a 0.1 \msun\ object in a circular orbit with a radius of 2 au would result in a RV semi-amplitude of 3.0 \kms\ (with a period of 4.0 years), and a 10 \mjup\ object would result in a RV semi-amplitude of 310 \ms\ (with a period of 4.5 years). While this can be used to rule out the existence of a stellar companion within 0.1 arcsec of the target, we note that these RV semi-amplitudes assume an edge-on orbit (orbital inclination of 90 degrees). A face-on orbit can in principle be undetected by the HARPS RVs. Gaia time series astrometry expected to be published within the next few years should be sensitive to massive companions regardless of orbital orientation.

\section{Investigating the orbital RV signal}
\label{sec:orbit}

\begin{figure}
\includegraphics[width=3.6in]{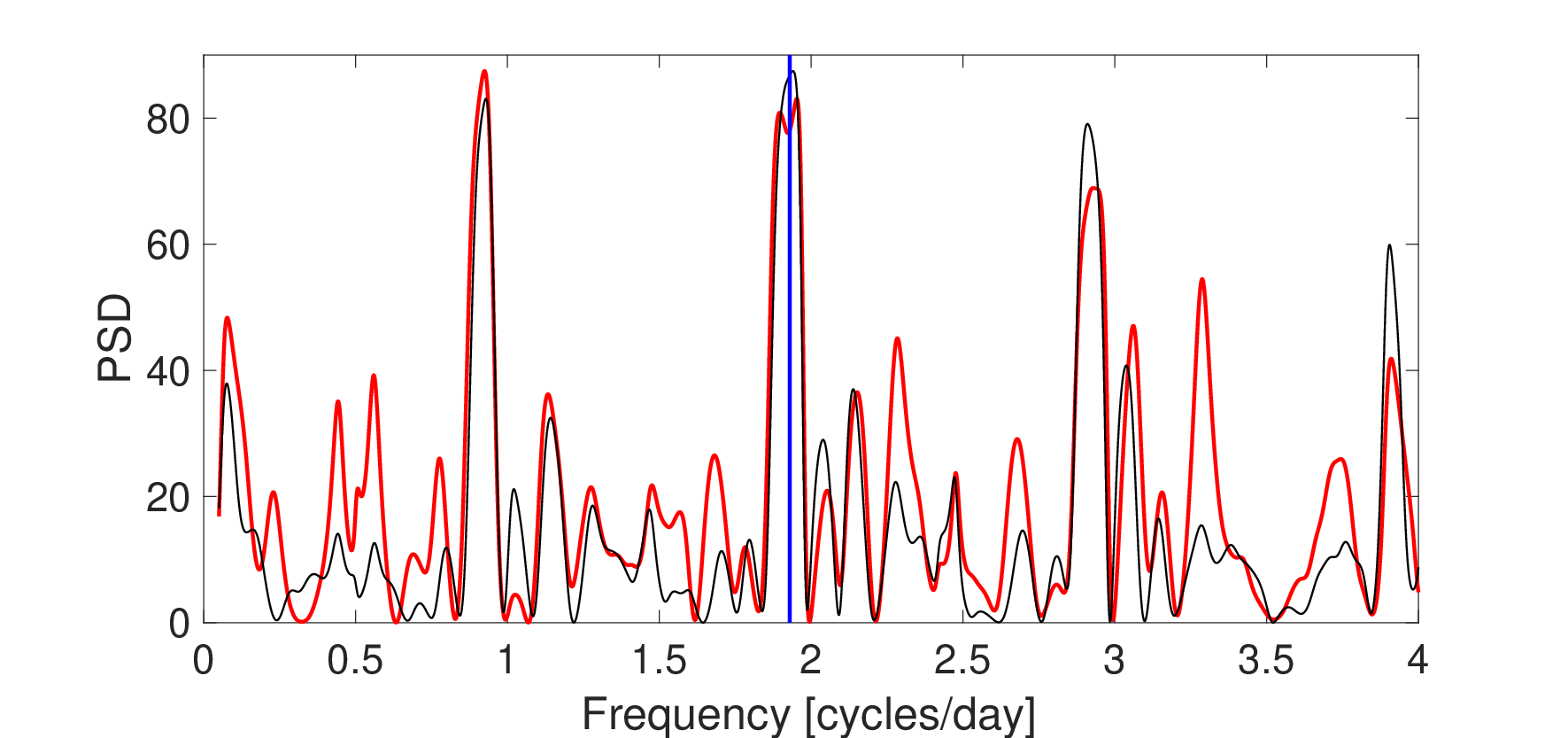}
\caption{Lomb-Scargle power spectrum density (PSD) of the HARPS RVs after subtracting a linear trend, shown by a red line. In gray is the PSD of a pure sinusoidal signal with the transit period injected at the data time stamps, scaled to the same level as the original PSD to allow visual comparison. The vertical blue line marks the transit period. 
}
\label{fig:lsper}
\end{figure}

\begin{figure}
\includegraphics[width=3.3in]{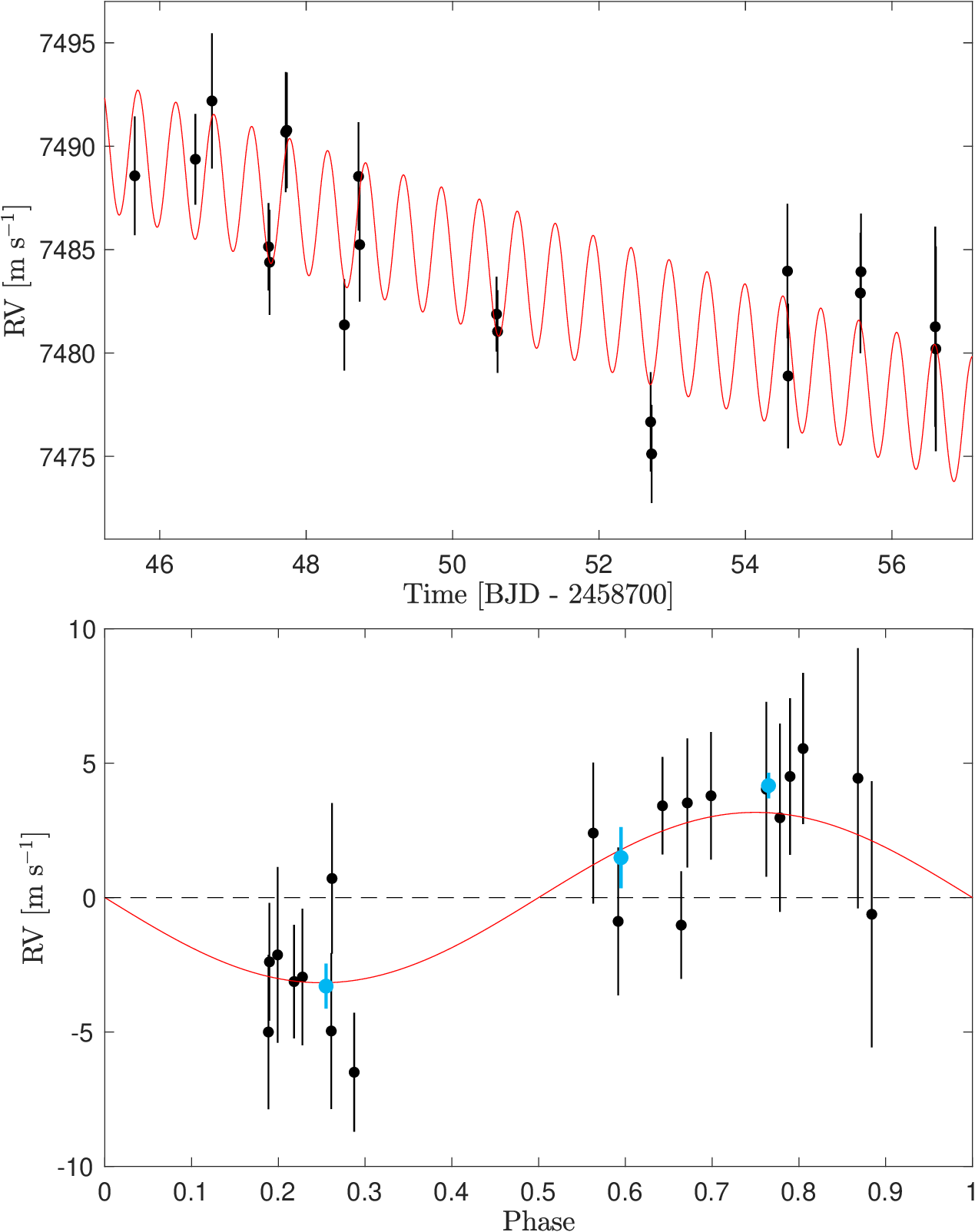}
\caption{{\it Top}: HARPS RVs as a function of time (black), overplotted with the fitted orbit model in red. The model includes a linear trend and a circular orbit. {\it Bottom:} Phase folded RVs (black) after subtracting the linear trend and the RV zero point, overplotted by the fitted circular orbit (red). The binned RV curve is marked in light blue, and a dashed horizontal black line is plotted at zero RV for reference. The analysis presented in this figure includes only the 20 HARPS RVs obtained in 2019.
}
\label{fig:rvmodel}
\end{figure}

We investigated the HARPS RVs to see if they reveal the host star's orbital motion. We have done that in two ways. We first looked for a periodic signal in the RV data set (see \figr{lsper}), and in a second analysis we fitted a circular orbit while fixing the period and phase to be consistent with the transit ephemeris (see \figr{rvmodel}). In both analyses we used from \tabr{rvs} the 20 RVs obtained during 2019 while ignoring the 5 RVs from the HARPS archival RVs. The reason for doing so is that the RV zero point offset following the 2015 HARPS fiber replacement is at the level of a few \ms\ \citep{locurto15}, comparable to the expected RV semi-amplitude. And, the 5 RVs obtained before 2015 are insufficient for a reliable determination of the pre-2015 RV zero point. 

Our first analysis was aimed at looking for a periodic signal in the HARPS RVs. As the RVs show a linear trend (see \figr{rvmodel}), we first fitted and removed a linear trend. The fitted slope was $-0.950 \pm 0.036$ \msd. We then performed a Lomb-Scargle period analysis \citep{lomb76, scargle82} of the detrended RVs, shown in \figr{lsper}. The periodogram shows a strong peak at the transit frequency, while showing a few other strong peaks. Those other peaks are due to aliasing, since the periodogram of a pure sinusoidal signal at the orbital period injected at the RV time stamps shows a highly similar structure. The periodogram of that injected signal is also shown in \figr{lsper} in gray, scaled to the original periodogram for visibility. The two periodograms show similar peaks, and specifically the three strongest peaks, including the one at the transit period, are almost identical. We conclude that the RVs show a periodic sinusoidal variability at the transit period.

In our second analysis we used \texttt{radvel} \citep{fulton18} to fit for a circular orbit with a linear trend.  We adopted the RV slope reference epoch at BJD of 2,458,751, at the center of the time period covered by RVs. We fitted for the orbital semi-amplitude $K$, the RV zero point $\gamma$, a linear trend $\dot{\gamma}$, and RV jitter $\sigma_{RV}$. We used a Gaussian prior on the period and transit time following the results of the light curve fit (\tabr{params}). The fitted parameters are listed in \tabr{params} and the fitted model is plotted in \figr{rvmodel}. The fitted RV trend is -1.13 $\pm$ 0.24 \msd, consistent with the trend fitted above in our first analysis. The fitted RV semi-amplitude was $K = 3.17 \pm 0.85$ \ms, representing a detection close to \sig{4} significance. We tested our analysis by rerunning it while setting the jitter term equal to zero. That analysis resulted in identical fitted parameters. We also attempted to fit an eccentric orbit (with and without an RV jitter term), but the resulting eccentricity was so poorly constrained that we opted to fit only for a circular orbit, which is expected for this short orbital period system. 

Using the measured RV semi-amplitude ($3.17 \pm 0.85$ \ms), the host star mass ($0.381 \pm 0.019$ \msun), and the orbital period we estimated the planet mass to be $M_p = 2.09 \pm 0.56$ \mearth. We acknowledge that this mass measurement has a relatively low statistical significance (\sig{\approx4}), and deriving it includes fitting a linear RV trend whose nature is currently not known. Therefore, we caution that the true uncertainty in the mass may be larger than 0.56 \mearth.

\section{Discussion}
\label{sec:dis}

\planet\ joins a small but growing group of small planets orbiting nearby M dwarf stars (e.g., \citealt{charbonneau09, gillon16, vanderspek19, kostov19, gunther19b, gan20}). It also joins the group of small planets orbiting at very short periods, commonly called ultra short periods, or USPs \citep[e.g.,][]{leger09, batalha11, sanchis14, adams16, winn18}. USPs orbital period ranges from about one day down to less than 10 hours \citep[e.g.,][]{howard13, dai17}, and even as short as $\approx$4 hours, especially around M dwarfs \citep{ofir13, rappaport13, smith18}. Planets in this group tend to be smaller than 2 \rearth\ \citep{winn18}, and are believed to have undergone photo-evaporation which removed their atmosphere \citep{lundkvist16, owen17}. With  a radius of $1.193 \pm 0.074$ \rearth\ and an equilibrium temperature of $1089 \pm 69$ K, it is likely that \planet\ have also went through that process.

As shown in \figr{raddist}, \target\ is one of the closest planet host stars to the Sun to host a planet with a measured radius. This proximity allows probing the system for massive planets at wide orbits with Gaia astrometric data that will be published in the next few years. Given \target's distance and mass, the astrometric signal of an orbiting planet is $123 (M_p/\mjup)(a/{\rm au})\ \mu$as \citep[e.g.,][]{perryman14, sozzetti14}, compared to an expected astrometric precision\footnote{\url{https://www.cosmos.esa.int/web/gaia/science-performance}} of $\approx10\ \mu$as for an M type star as bright as \target. Therefore, while the Gaia astrometric data will not be sensitive to \planet\ it will complement and extend the ongoing RV monitoring in identifying additional companions. For example, the full HARPS RV time series, including archival RVs, suggests \target\ is a single star. The Gaia astrometric data will allow confirming that independently of orbital inclination.

What is the origin of the long-term trend seen in the HARPS RV data? One possibility is that it is induced by another planet in the system, with an orbital period a few times longer than the 12 day time span of the RV data. This is supported by the fact that many USPs reside in multi-planet systems \citep{winn18}. Another possibility is that the trend is caused by stellar activity. To check that possibility we examined the 20 HARPS spectra cross correlation function (CCF), derived by cross correlating the spectra with a synthetic M-dwarf template spectrum. We calculated the correlation between the RVs (CCF center) and the CCFs FWHM, contrast, and bisector span, listed in \tabr{rvs}. We also calculated the correlation between the RVs and the activity index $S$ derived from the HARPS spectra. In all cases the correlation was below 0.20 in absolute value. In comparison, the width (standard deviation) of a distribution of 10$^5$ cross correlation values of the RVs with randomly permuted vectors of the parameters listed above is 0.23, indicating that the correlations are not statistically significant. This is consistent with our findings in \secr{star},  that the star is expected to show only low stellar activity. Future RVs will result in an improved dynamical estimate of the planet's mass, and will also allow for a better investigation of the nature of the RV trend.

\begin{figure}
\includegraphics[width=3.3in]{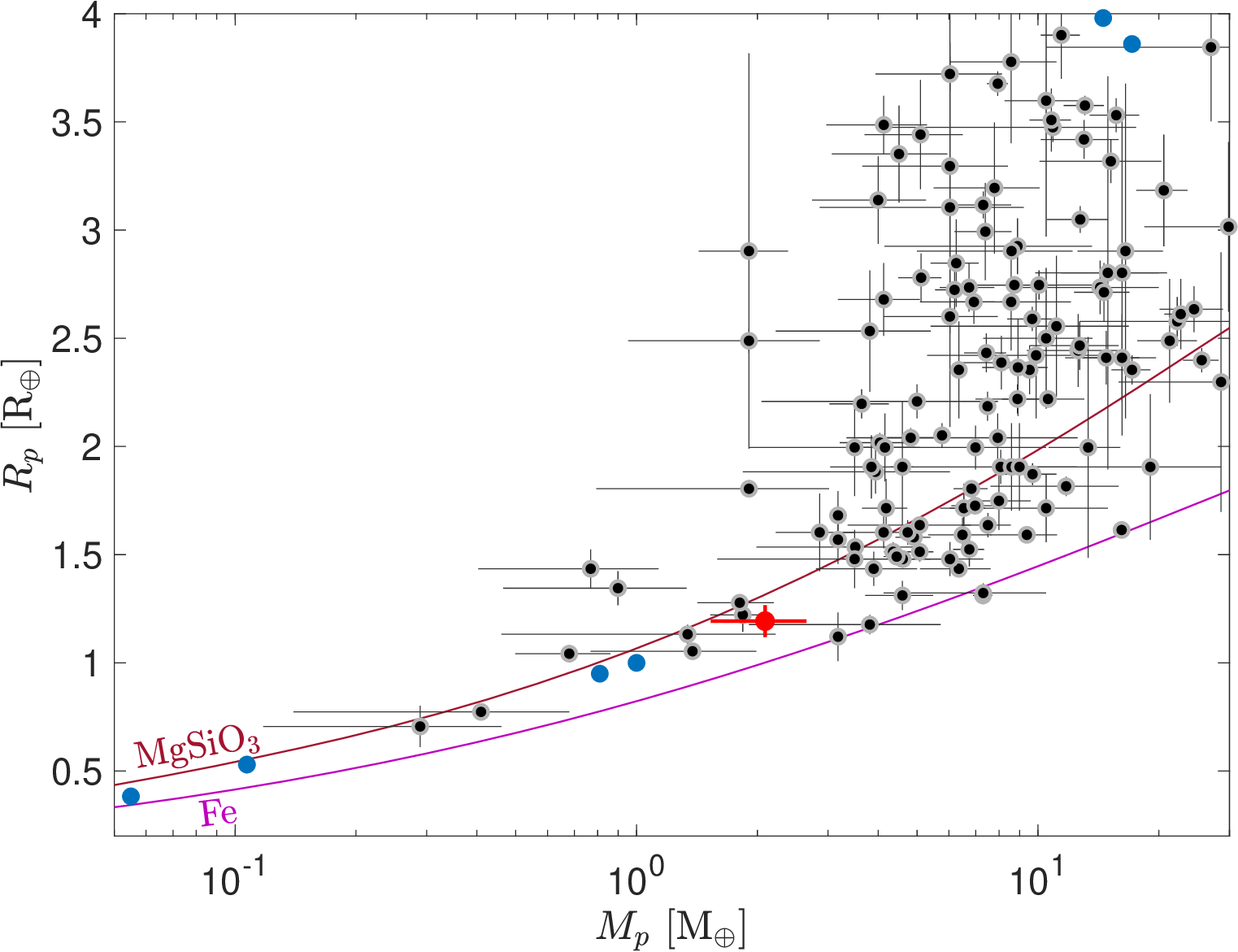}
\caption{Planet radius vs.~planet mass with \planet\ marked in red. \planet\ is marked in red, and models of pure iron (Fe) and pure rock (MgSiO$_{\rm 3}$) compositions are overplotted, taken from \citet{zeng16}. The plot omits planets with a poor radius or mass measurement where the measured parameter is smaller than 1.5 times the uncertainty. Solar system planets are marked in blue. Data taken from the NASA Exoplanet Archive on 2019 Oct 16.
}
\label{fig:radmass}
\end{figure}

\begin{figure}
\includegraphics[width=3.6in]{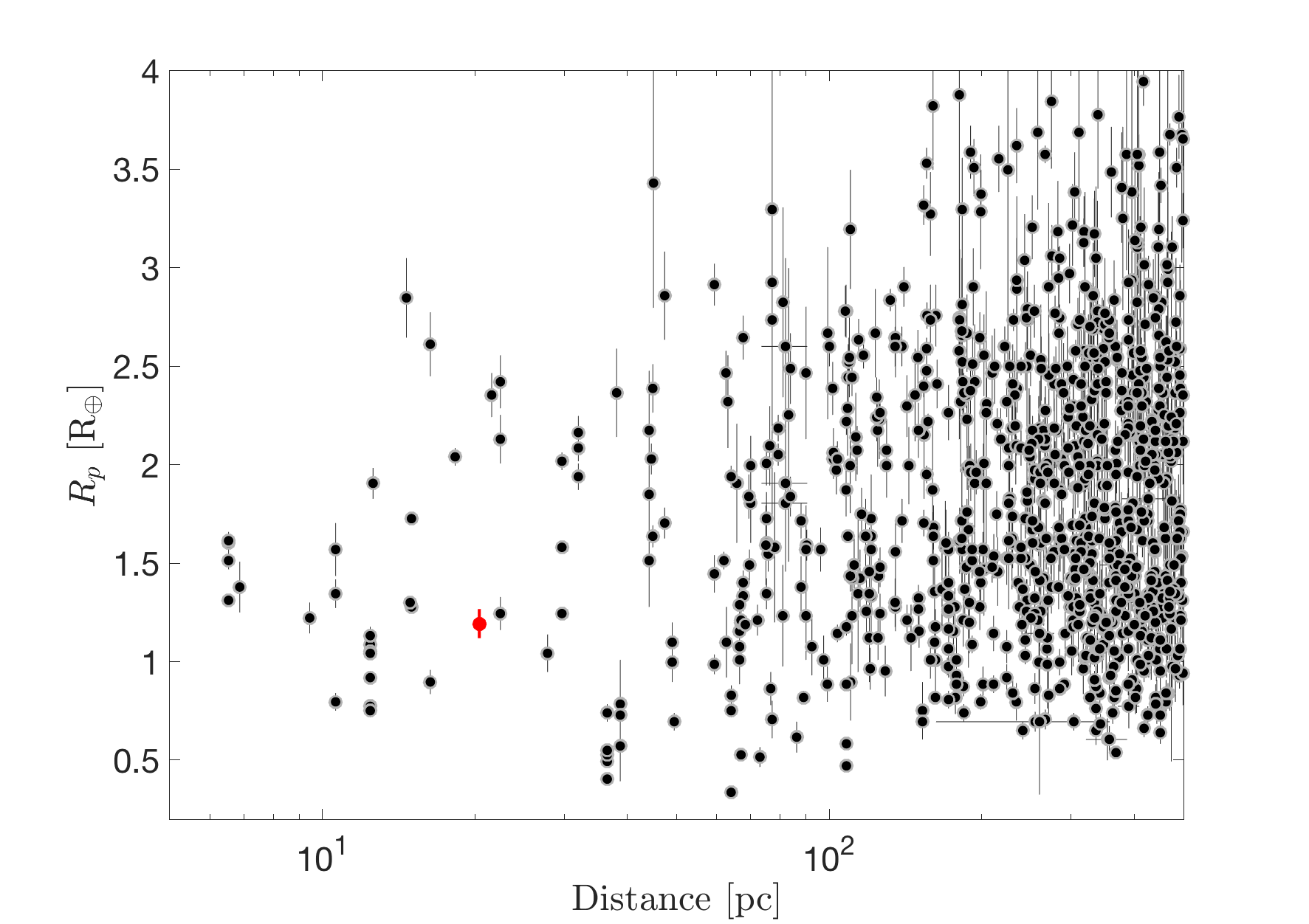}
\caption{Planet radius as a function of host star distance for planets with known radius. \planet\ is marked in red. The plot does not include planets with a poor radius measurement where radius is smaller than 1.5 times the radius uncertainty. We note that unlike \figr{radmass} this plot includes planets without a mass measurement. Data taken from the NASA Exoplanet Archive on October 16, 2019.
}
\label{fig:raddist}
\end{figure}

While our mass measurement is marginal, it is interesting to examine the position of \planet\ in the radius-mass diagram, plotted in \figr{radmass}. The plot shows that \planet\ is among the smallest planets with an estimated mass. Its period and radius place it below the gap in the distribution of close-in planetary radii around M dwarfs \citep{cloutier19}, a gap that is at a slightly smaller radius range than for planets orbiting Sun-like stars \citep{fulton17, cloutier19}. It was shown by several authors that the majority of small planets, below the radius gap, with measured radii and masses have terrestrial bulk compositions  \citep{dai19, jontof19, otegi19}. This prediction is consistent with \planet's measured bulk density, of $6.8 \pm 2.2$ g cm$^{-3}$, and its position between theoretical radius-mass relations assuming pure iron and pure rock compositions \citep{zeng16}.


\target's brightness ($V$ = 12.19 mag, $K$ = 7.92 mag) and the short orbital period (0.518 day, or 12.4 hours) make it a potential target for transmission and emission spectroscopy, which can reveal whether or not the planet has an atmosphere. Following \citet{kempton18} the system's transmission spectroscopy metric (TSM) is $20.0\pm7.0$, and its emission spectroscopy metric (ESM) is $16.4\pm3.5$. Therefore, while both metrics have significant uncertainties, the prospects for transmission spectroscopy are not promising, although the prospects for emission spectroscopy are. For example, at 10 \mic\ the expected secondary eclipse depth is about 150 ppm. Therefore, \planet\ is a potential target for IR secondary eclipse and phase curve measurements, which in turn can probe the planet's atmosphere, or its absence \citep{kreidberg19}. Additional RVs and \tess\ photometry will lead to improved planet mass and radius measurements, in turn leading to more precise TSM and ESM. Additional RVs are currently being gathered, and additional \tess\ photometry is scheduled to be obtained during the first sector of the \tess\ extended mission, Sector 27, in July 2020.

\section{Summary}
\label{sec:sum}

\planet\ joins the short but growing list of small planets orbiting bright and nearby stars discovered by \tess\ that are amenable to detailed characterization.

We took advantage of the star's properties, specifically its small size and high proper motion, to validate the transit signal detected in \tess\ data as originating from a star-planet system.
We also obtained a marginal planet mass measurement, and ongoing RV monitoring will allow an improved mass estimate. If successful it will lead to a precise planet mass measurement below the radius gap \citep{fulton17, cloutier19}. Long term RV monitoring will also allow looking for other planets in the system, as will future \tess\ photometry to be obtained during the \tess\ extended mission. 

The host star proximity and brightness and the short orbital period make this star-planet system an attractive target for detailed characterization. These investigations include studying the planet's atmosphere \citep{kempton18}, and using future Gaia astrometric data, combined with long term RV monitoring, to look for any currently unknown star, brown dwarf, or massive planet orbiting the host star.

\acknowledgments

M.N.G.~and C.X.H.~acknowledge support from MIT's Kavli Institute as Torres postdoctoral fellows.
T.D.~acknowledges support from MIT’s Kavli Institute as a Kavli postdoctoral fellow.
N.A.-D.~acknowledges the support of FONDECYT project 3180063.
C.D.D.~was supported by the NASA TESS Guest Investigator Program via grant 80NSSC18K1583.
K.H.~acknowledges support from the UK STFC grant ST/R000824/1.
J.N.W thanks the Heising-Simons Foundation for its support.
We acknowledge the use of \tess\ Alert data. These data are derived from pipelines at the \tess\ Science Office and at the \tess\ Science Processing Operations Center.
Funding for the \tess\ mission is provided by NASA's Science Mission directorate.
Based on observations collected at the European Organisation for Astronomical Research in the Southern Hemisphere under ESO programme 0103.C-0449(A).
This paper includes data collected by the \tess\ mission, which are publicly available from the Mikulski Archive for Space Telescopes (MAST).
Resources supporting this work were provided by the NASA High-End Computing (HEC) Program through the NASA Advanced Supercomputing (NAS) Division at Ames Research Center [for the production of the SPOC data products].
This research has made use of the NASA Exoplanet Archive, which is operated by the California Institute of Technology, under contract with NASA under the Exoplanet Exploration Program.
The Digitized Sky Surveys were produced at the Space Telescope Science Institute under U.S.~Government grant NAG W-2166. The images of these surveys are based on photographic data obtained using the Oschin Schmidt Telescope on Palomar Mountain and the UK Schmidt Telescope. The plates were processed into the present compressed digital form with the permission of these institutions.
This work makes use of observations from the LCOGT network.
The MEarth Team gratefully acknowledges funding from the David and Lucile Packard Fellowship for Science and Engineering (awarded to D.C.). This material is based upon work supported by the National Science Foundation under grants AST-0807690, AST-1109468, AST-1004488 (Alan T.~Waterman Award), and AST-1616624. 
This work is made possible by a grant from the John Templeton Foundation, which also supports J.G.W. The opinions expressed in this publication are those of the authors and do not necessarily reflect the views of the John Templeton Foundation.\\
{\it Facilities:} 
\facility{\tess},
\facility{ESO:3.6m (HARPS)},
\facility{VLT:Antu (NaCo)},
\facility{MEarth},
\facility{WASP-South},
\facility{LCO:1m (Sinistro)},
\facility{LCO:2m (Spectral)}

{\it Software:} 
\texttt{radvel} \citep{fulton18},
\texttt{allesfitter} \citep{gunther19a},
{\tt AstroImageJ} \citep{collins17}




\end{document}